\documentstyle[preprint]{ptptex}
\title{%        %You can use \\ for explicit line-break
On the Boson Number Operator in the Deformed Boson Scheme
}
\author{%       %Use \sc for the family name
Atsushi {\sc Kuriyama},$^{1}$ 
Constan\c{c}a {\sc Provid\^encia}$^{2}$, \\
Jo\~ao da {\sc Provid\^encia}$^{2}$, Yasuhiko {\sc Tsue}$^{3}$ 
and Masatoshi {\sc Yamamura}$^{1}$
}

\inst{%         %Affiliation, neglected when [addenda] or [errata]
$^1$Faculty of Engineering, Kansai University, Suita 564-8680, Japan\\
$^{2}$Departamento de Fisica, Universidade de Coimbra, P-3000 Coimbra, 
Portugal\\
$^{3}$Physics Division, Faculty of Science, Kochi University, Kochi 780-8520, 
Japan
}

\recdate{%      %Editorial Office will fill in this.
}

\abst{%       %this abstract is neglected when [addenda] or [errata]
Concerning the energy of harmonic oscillator, a prescription is proposed 
for making the original form unchanged even after $q$-deformation. 
Applicability of the prescription is limited, but, it can be applied 
to various cases which are well known. 
}

\begin{document}

\maketitle

%\section{Introduction}

An idea of $q$-deformation serves us an interesting viewpoint for the 
understanding of dynamics of many-body systems described by boson operators. 
For the $q$-deformation, the present authors recently gave a possible scheme 
in relation to the generalized boson coherent state.\cite{1,2,3}
It is, in some sense, an extension of the idea given by 
Penson and Solomon.\cite{4} 
Before entering the central part of this paper, first, we recapitulate 
the basic idea of Ref. \citen{1}.

We treat a boson system described by one kind of boson 
$({\hat c} , {\hat c}^*)$. As the deformation of $({\hat c}, {\hat c}^*)$, 
the operator $({\hat \gamma} , {\hat \gamma}^*)$ is introduced in the 
definition 
\begin{equation}\label{1}
{\hat \gamma}=F({\hat N}){\hat c} \ , \qquad 
{\hat \gamma}^*={\hat c}^* F({\hat N}) \ .
\end{equation}
Here, $F({\hat N})$ denotes a function of the boson number operator, 
${\hat N}$, defined by 
\begin{equation}\label{2}
{\hat N}={\hat c}^*{\hat c} \ .
\end{equation}
The function $F({\hat N})$ obeys a certain condition which is proper 
to each case under investigation. 
The present deformation is characterized by the function $F({\hat N})$. 
The commutation relations $[{\hat c} , {\hat c}^*]$ and 
$[{\hat \gamma} , {\hat \gamma}^*]$ are as follows : 
\begin{subequations}\label{3}
\begin{eqnarray}
& &[{\hat c} , {\hat c}^* ]=1 \ , 
\label{3a}\\
& &[{\hat \gamma} , {\hat \gamma}^* ]=({\hat N}+1)F({\hat N})^2
-{\hat N}F({\hat N}-1)^2 \ . 
\label{3b}
\end{eqnarray}
\end{subequations}
Later, we discuss the case in which the relation (\ref{3b}) does not hold. 
If following our deformed boson scheme, any operator which is expressed 
in terms of a function $({\hat c} , {\hat c}^*)$, 
${\hat O}=O({\hat c}, {\hat c}^*)$, is deformed by replacing 
$({\hat c}, {\hat c}^*)$ with $({\hat \gamma} , {\hat \gamma}^*)$ : 
\begin{equation}\label{4}
{\hat O}_q=O({\hat \gamma}, {\hat \gamma}^*) \ .
\end{equation}
However, in the case of the Hamiltonian, a special consideration is 
necessary for the replacement (\ref{4}).

Usually, the Hamiltonian under investigation, ${\hat H}$, consists of 
two terms : 
\begin{equation}\label{5}
{\hat H}={\hat H}^{(0)}+{\hat H}^{(1)} \ , 
\end{equation}
\setcounter{equation}{4}
\vspace{-0.8cm}
\begin{subequations}
\begin{eqnarray}
& &{\hat H}^{(0)}=\omega {\hat c}^*{\hat c} \ , 
\label{5a}\\
& &{\hat H}^{(1)}=V({\hat c}, {\hat c}^*) \ . 
\label{5b}
\end{eqnarray}
\end{subequations}
Here, ${\hat H}^{(0)}$ denotes energy of the harmonic oscillator with the 
frequency $\omega$. On the other hand, the term ${\hat H}^{(1)}$ makes 
the system under the investigation deviate from the harmonic oscillator. 
With the aid of the $q$-deformation, we are able to obtain various terms 
which make the system deviate from the harmonic oscillator. 
The formal application of the form (\ref{4}) to the Hamiltonian (\ref{5}) 
gives us 
\begin{equation}\label{6}
{\hat H}_q={\hat H}_q^{(0)}+{\hat H}_q^{(1)} \ , \qquad\qquad\qquad\qquad\ \ 
\end{equation}
\vspace{-1cm}
\setcounter{equation}{5}
\begin{subequations}
\begin{eqnarray}
& &{\hat H}_q^{(0)}=\omega{\hat \gamma}^*{\hat \gamma}
=\omega{\hat N}F({\hat N}-1)^2 \ , 
\label{6a}\\
& &{\hat H}_q^{(1)}=V({\hat \gamma}, {\hat \gamma}^*)
=V(F({\hat N}){\hat c} , {\hat c}^*F({\hat N})) \ . 
\label{6b}
\end{eqnarray}
\end{subequations}
Certainly, we can expect that ${\hat H}_q^{(1)}$ makes the system deviate 
from the harmonic oscillator in various forms and, further, we see 
that ${\hat H}_q^{(0)}$ contains also the part which plays the same 
role as that of ${\hat H}_q^{(1)}$. However, if we insist on the 
original picture which the Hamiltonian tells us, ${\hat H}_q^{(0)}$ 
should be of the form of the harmonic oscillator with the frequency 
$\omega$. This means that ${\hat H}_q^{(0)}$ is at most of the form 
\begin{equation}\label{7}
{\hat H}_q^{(0)}=E_0+\omega {\hat c}^*{\hat c} \ . 
\end{equation}
Here, $E_0$ denotes constant. In order to obtain the form (\ref{7}) in 
the case of the $su(2)$- and the $su(1,1)$-algebraic models, we needed 
special device for each case which is shown in Refs.\citen{2} and 
\citen{3}.

Main aim of the present note is to give a prescription which leads us 
to the form (\ref{7}) automatically. 
First, we introduce the following operator : 
\begin{equation}\label{8}
{\hat M}={\hat c}^*{\hat c}+G({\hat b})-G(0) \ . 
\end{equation}
Here, ${\hat b}$ is defined as 
\begin{equation}\label{9}
{\hat b}=(1/2)\cdot ([{\hat c} , {\hat c}^* ]-1) \ . 
\end{equation}
The second and the third terms on the right-hand side of the relation 
(\ref{8}), $G({\hat b})$ and $G(0)$, denote the values of any function 
$G(x)$ at $x={\hat b}$ and $0$, respectively. 
Since ${\hat b}$ is null operator, we can see that the operator 
${\hat M}$ is nothing but the boson number operator ${\hat N}$. 
Formal replacement (\ref{4}) for the relation (\ref{8}) leads us to 
\begin{equation}\label{10}
{\hat M}_q={\hat \gamma}^*{\hat \gamma}+G({\hat \beta})-G(0) \ . 
\end{equation}
Here, ${\hat \beta}$ is given by 
\begin{equation}\label{11}
{\hat \beta}=(1/2)\cdot ([{\hat \gamma} , {\hat \gamma}^*]-1) \ . 
\end{equation}
As can be seen in the relation (\ref{3b}), ${\hat \beta}$ is not null 
operator, i.e., 
$G({\hat \beta})\neq G(0)$. Substituting the relation (\ref{3b}) into 
the definition (\ref{11}), ${\hat \beta}$ is given as 
\setcounter{equation}{10}
\begin{subequations}
\begin{equation}\label{11a}
{\hat \beta}=(1/2)\cdot [({\hat N}+1)F({\hat N})^2
-{\hat N}F({\hat N}-1)^2-1] \ .
\end{equation}
\end{subequations}
If the relation (\ref{11a}) can be inversely solved, ${\hat N}$ can be 
expressed in terms of the function of ${\hat \beta}$ and we denote it as 
\begin{equation}\label{12}
{\hat N}=g({\hat \beta}) \ .
\end{equation}

As is clear from the form (\ref{10}), ${\hat M}_q$ depends on the 
function $G(x)$. In order to fix the form of ${\hat M}_q$ which is 
consistent to the form (\ref{7}), we require the following condition~ : 
\begin{equation}\label{13}
{\hat \gamma}^*{\hat \gamma}+G({\hat \beta})
\ (={\hat N}F({\hat N}-1)^2+G({\hat \beta}))\ ={\hat N} \ .
\end{equation}
Then, ${\hat M}_q$ can be expressed as 
\begin{equation}\label{14}
{\hat M}_q={\hat c}^*{\hat c}-G(0) \ .
\end{equation}
Then, the quantity $E_0$ in the relation (\ref{7}) is given by 
\begin{equation}\label{15}
E_0=-\omega G(0) \ . 
\end{equation}
The relations (\ref{12}) and (\ref{13}) determine the form of 
$G({\hat \beta})$ : 
\begin{eqnarray}
& &G({\hat \beta})={\hat N}-{\hat N}F({\hat N}-1)^2
=g({\hat \beta})[1-F(g({\hat \beta})-1)^2] \ , 
\label{16}
\end{eqnarray}
\vspace{-0.5cm}
\setcounter{equation}{15}
\begin{subequations}
\begin{equation}
G(0)=g(0)[1-F(g(0)-1)^2] \ . \qquad\qquad\qquad\qquad\quad\ 
\end{equation}
\end{subequations}
Thus, ${\hat M}$ and ${\hat M}_q$ are expressed in the form 
\begin{eqnarray}
& &{\hat M}={\hat c}^*{\hat c}+g({\hat b})[1-F(g({\hat b})-1)^2]
-g(0)[1-F(g(0)-1)^2] \ , 
\label{17}\\
& &{\hat M}_q={\hat c}^*{\hat c}-g(0)[1-F(g(0)-1)^2] \ . 
\label{18}
\end{eqnarray}

Let us show two examples. Example (i) is given by 
$F({\hat N})=\sqrt{1\mp {\hat N}/n_0}$. 
The upper and the lower sign correspond to the deformations to the 
$su(2)$- and $su(1,1)$-algebras in the Holstein-Primakoff 
representation, respectively.\cite{2} 
In this case, the relation (\ref{11a}) gives us 
\begin{equation}\label{19}
{\hat \beta}=\mp {\hat N}/n_0 \ , \quad
{\rm i.e.,}\quad g({\hat \beta})=\mp n_0{\hat \beta} \ . 
\end{equation}
Then, ${\hat M}$ and ${\hat M}_q$ shown in the forms (\ref{17}) 
and ({\ref{18}) can be expressed as 
\begin{eqnarray}
& &{\hat M}={\hat c}^*{\hat c}+{\hat b}(1\pm n_0{\hat b}) \ , 
\label{20}\\
& &{\hat M}_q={\hat c}^*{\hat c} \ . 
\label{21}
\end{eqnarray}
Example (ii) is related to $F({\hat N})=\sqrt{{\hat N}/n_0-1}$. 
This case produces the second Holstein-Primakoff representation of the 
$su(1,1)$-algebra\cite{2} and we have 
\begin{equation}\label{22}
{\hat \beta}={\hat N}/n_0-1 \ , \quad {\rm i.e.,}\quad
g({\hat \beta})=n_0({\hat \beta}+1) \ . 
\end{equation}
Then, ${\hat M}$ and ${\hat M}_q$ shown in the forms (\ref{17}) 
and (\ref{18}) can be expressed as 
\begin{eqnarray}
& &{\hat M}={\hat c}^*{\hat c}+{\hat b}(1-n_0{\hat b}) \ , 
\label{23}\\
& &{\hat M}_q={\hat c}^*{\hat c}-(n_0+1) \ . 
\label{24}
\end{eqnarray}
The results (\ref{21}) and (\ref{24}) are natural. In the case of the 
example (i), $N$ can run in the region $0, 1, 2, \cdots$, because of the 
positive-definiteness of $(1\mp {\hat N}/n_0)$. 
On the other hand, the example (ii) tells us that in this case $N$ can 
run in the region $(n_0+1), (n_0+2), (n_0+3), \cdots$, because of the 
positive-definiteness of $({\hat N}/n_0-1)$. 
The results (\ref{21}) and (\ref{24}) support this fact.

However, formal application of the above procedure cannot be directly 
permitted, for example, for the case 
\begin{equation}\label{25}
F({\hat N})=\phi({\hat N})/\sqrt{{\hat N}+1} \ .
\end{equation}
Here, $\phi(n)$ is assumed to be well-behaved function and 
$\phi(-1)=0$. For example, we have the following form : 
\begin{equation}\label{26}
\phi(n)=\sqrt{[1-q^{-2(n+1)}]/[1-q^{-2}]} \ .
\end{equation}
This form is a possible modification of the most popular form in the 
$q$-deformation, \break
$\sqrt{[q^{n+1}-q^{-(n+1)}]/[q-q^{-1}]}$, and 
the form adopted in Ref.\citen{4}, $q^{n/2}$.\cite{1} 
Here, $q$ denotes a positive parameter. Formal application of the form 
(\ref{25}) to the relation (\ref{11a}) results the appearance of the 
operator ${\hat N}/{\hat N}$, which cannot be defined. 
In this case, the relation (\ref{11a}) should be rewritten as 
\begin{eqnarray}\label{27}
{\hat \beta}&=&
(1/2)\cdot [({\hat N}+1)F({\hat N})^2-{\hat Q}_0{\hat N}F({\hat N}-1)^2-1] 
\nonumber\\
&=&(1/2)\cdot[\phi({\hat N})^2-{\hat Q}_0\phi({\hat N}-1)^2-1] \ .
\end{eqnarray}
Here, ${\hat Q}_0$ denotes the projection operator which rejects the 
vacuum state for the boson $({\hat c}, {\hat c}^*)$. 
Further, the relation (\ref{16}) should be rewritten as 
\begin{equation}\label{28}
G({\hat \beta})={\hat N}-{\hat Q}_0{\hat N}
F({\hat N}-1)^2
={\hat N}-{\hat Q}_0\phi({\hat N}-1)^2 \ . 
\end{equation}
Since there exists the condition $\phi(-1)=0$, we have the relation 
${\hat Q}_0\phi({\hat N}-1)^2=\phi({\hat N}-1)^2$ and, then, 
the relations (\ref{27}) and (\ref{28}) are reduced to the 
following form~ : 
\setcounter{equation}{26}
\begin{subequations}
\begin{equation}\label{27a}
{\hat \beta}=(1/2)\cdot[\phi({\hat N})^2-\phi({\hat N}-1)^2-1] \ , 
\end{equation}
\end{subequations}
\vspace{-0.5cm}
\begin{subequations}
\begin{equation}\label{28a}
G({\hat \beta})={\hat N}-\phi({\hat N}-1)^2 \ . \qquad\qquad\quad\ 
\end{equation}
\end{subequations}
In the same procedure as the case already shown, we can treat the relations 
(\ref{27a}), (\ref{17}) and (\ref{18}). For the case (\ref{26}), the 
following solution is obtained : 
\begin{eqnarray}
& &{\hat \beta}=-(1/2)\cdot(1-q^{-2{\hat N}}) \ , \quad{\rm i.e.,}\quad
g({\hat \beta})=-(1/2)\cdot \log(1+2{\hat \beta})/\log q \ , 
\label{29}\\
& &{\hat M}={\hat c}^*{\hat c}+2{\hat b}/(1-q^{-2})
-(1/2)\cdot \log(1+2{\hat b})/\log q \ , 
\label{30}\\
& &{\hat M}_q={\hat c}^*{\hat c} \ . 
\label{31}
\end{eqnarray}
In the case of $\phi(-1)\neq 0$, further device should be investigated. 
In this note, we will not contact with this problem.

We have one more example which should be carefully treated : 
\begin{equation}\label{32}
\phi(n)=1 \ .
\end{equation}
The relation (\ref{27}) in this case is written as 
\begin{equation}\label{33}
{\hat \beta}=-(1/2)\cdot{\hat Q}_0 \ . 
\end{equation}
The relation (\ref{33}) cannot be solved inversely in terms of 
${\hat N}=g({\hat \beta})$. Then, we note the relation 
\begin{equation}\label{34}
\lim_{\varepsilon\rightarrow 0}{\hat N}/({\hat N}+\varepsilon)={\hat Q}_0 \ .
\end{equation}
Replacing ${\hat Q}_0$ in the relations (\ref{27}) and (\ref{28}) 
with ${\hat N}/({\hat N}+\varepsilon)$ and regarding $\varepsilon$ as 
an infinitesimal parameter, we formulate the problem. Under this 
procedure, we have the following result : 
\begin{eqnarray}
& &{\hat \beta}=-(1/2)\cdot{\hat N}/({\hat N}+\varepsilon) \ , 
\quad{\rm i.e.,}\quad
g({\hat \beta})=-\varepsilon\cdot 2{\hat \beta}/(2{\hat \beta}+1) \ , 
\label{35}\\
& &{\hat M}={\hat c}^*{\hat c}+2{\hat b}
-\varepsilon\cdot 2{\hat b}/(2{\hat b}+1) \ , 
\label{36}\\
& &{\hat M}_q={\hat c}^*{\hat c} \ . 
\label{37}
\end{eqnarray}
%Of course, we used the relation 
Here, we used the relation 
\begin{equation}\label{38}
G({\hat \beta})=2{\hat \beta}-\varepsilon\cdot
2{\hat \beta}/(2{\hat \beta}+1) \ ,
%\ . 
\end{equation}
which can be derived from (\ref{28}), (\ref{12}) and (\ref{35}) carefully.

As was shown in the above, we could give a prescription, with the aid 
of which we can arrive at the aim mentioned in the introductory part 
of this note. Of course, the applicability is limited, however, 
it can be applied to various cases which we know. 
Then, our next interest becomes to investigate the deformation of 
${\hat H}^{(1)}$ which is given in the relation (\ref{5b}).

%\section*{Acknowledgements}
\vspace{0.2cm}

Two of the authors (Y. T. and M. Y.) would like to express their thanks to 
Professor J. da Provid\^encia, one of co-authors of this paper. 
He invited them to Coimbra in summer of 2002. During this stay, this work 
was initiated.

\end{document}